\documentclass[structabstract]{aa}  
\usepackage{natbib}
\bibpunct{(}{)}{;}{a}{}{,} 
\usepackage{graphicx}
\usepackage{txfonts}
\usepackage{lscape}
\newcommand{\xmm}{{\it XMM-Newton}}
\newcommand{\xmmnewton}{{\it XMM-Newton}}
\newcommand{\rosat}{{\it ROSAT}}

\newcommand{\chandra}{{\it Chandra}}
\newcommand{\fermi}{{\it Fermi}}

\newcommand{\ctb}{CTB~109}

\newcommand{\nh}{\mbox {$N_{\rm H}$}}

\newcommand{\mgxi}{Mg\,{\sc xi}}

\newcommand{\sixiii}{Si\,{\sc xiii}}

\newcommand{\sxv}{S\,{\sc xv}}
\newcommand{\about}{$\sim$\kern.03em}
\def\la{\mathrel{\hbox{\rlap{\hbox{\lower4pt\hbox{$\sim$}}}\hbox{$<$}}}}

\begin{document}

   \title{\chandra\ observation of the Galactic supernova remnant CTB\,109 
(G109.1--1.0)
         }

   \subtitle{}

   \titlerunning{\chandra\ observation of the SNR CTB\,109}

   \author{Manami Sasaki
          \inst{1}
          \and
          Paul P.\ Plucinsky\inst{2} 
          \and
          Terrance J.\ Gaetz\inst{2} 
          \and
          Fabrizio Bocchino\inst{3}
          }

   \institute{Institut f\"ur Astronomie und Astrophysik, 
              Universit\"at T\"ubingen,
              Sand 1, 
              D-72076 T\"ubingen, Germany,
              \email{sasaki@astro.uni-tuebingen.de}
         \and
              Harvard-Smithsonian Center for Astrophysics,
              60 Garden Street, Cambridge, MA 02138, USA
         \and
              INAF - Osservatorio Astronomico di Palermo, 
              Piazza del Parlamento 1, 
              90134, Palermo, Italy
             }

  \date{Received Dec., 3, 2012; accepted Feb., 7, 2013}

 
  \abstract
   {We study the X-ray emission of the Galactic supernova remnant (SNR) 
\ctb\ (G109.1--1.0), which is 
well-known for its enigmatic half-shell morphology both in radio and in X-rays
and is associated with the anomalous X-ray pulsar (AXP) 1E\,2259+586.}
   {We want to understand the origin of the X-ray bright feature inside the SNR called
the Lobe and the details of the interaction of the SNR shock wave with the ambient
interstellar medium (ISM).}
   {The Lobe and the northeastern part of the SNR were observed with \chandra\ ACIS-I.
We analysed the spectrum of the X-ray emission by dividing the entire observed emission
into small regions. The X-ray emission is best reproduced with one-component or
two-component non-equilibrium ionisation models depending on the position. 
In the two-component model, one emission component represents the shocked ISM and the
other the shocked ejecta.}
   {We detect enhanced element abundances, in particular for Si and Fe, in and 
around the Lobe. 
There is one particular region next to the Lobe with a high Si abundance
of 3.3 (2.6 -- 4.0) times the solar value. This is the first, unequivocal 
detection of ejecta in \ctb.
} 
   {The new \chandra\ data confirm that the Lobe was created by the interaction of
the SNR shock and the supernova ejecta with a dense and inhomogeneous medium in the 
environment of SNR \ctb. The newly calculated age of the SNR is 
$t \approx\ 1.4 \times 10^4$~yr.}


   \maketitle
%

\section{Introduction}

The Galactic supernova remnant \ctb\ (G109.1--1.0) 
is the host of the anomalous X-ray pulsar (AXP) 1E\,2259+586 
\citep{1981Natur.293..202F} and represents one of the most exotic 
and interesting objects in the X-ray sky.
The study of an SNR associated with an AXP provides valuable 
information on the environment in which it formed and an independent
estimate of the age of the objects. Given that there are only three firm
associations of SNRs with AXPs \citep{2008A&ARv..15..225M}, detailed 
studies of each association will produce progress in our understanding 
not only of SNRs but also of the AXPs.

The SNR \ctb\ has a
spectacular semi-circular morphology in both the X-ray and the radio. 
At a distance of 3.2$\pm$0.2~kpc \citep{2012ApJ...746L...4K}, it is
located next to a giant molecular cloud (GMC) complex and is 
one of the most 
striking examples of an interaction of an SNR with a molecular cloud. 
Since there is neither X-ray \citep[see Fig.\ 1 in][]{2004ApJ...617..322S} 
nor radio emission \citep{1981ApJ...246L.127H} where the western part of 
the shell would be 
expected, the semi-circular morphology of \ctb\ implies that the shock has
been greatly impeded or even stopped entirely by the GMC complex in the west.
A linear feature in CO (`CO arm') extends from the GMC complex to the local 
X-ray minimum in the northern half of the SNR \citep{1987A&A...184..279T}. 
Therefore, at least a part of the GMC complex extends in front of the 
remnant. \ctb\ has an X-ray bright interior region known as the `Lobe'. 
The Lobe is brighter than any part of the shell. Although it has been 
suggested that this feature is a jet associated with the 
AXP 1E\,2259+586 \citep{1983IAUS..101..429G},  
high-resolution images from \rosat\/ HRI \citep{1995MNRAS.277..549H},
\chandra\ \citep{2001ApJ...563L..45P}, as well as \xmmnewton\ 
\citep{2004ApJ...617..322S}, show no morphological connection with the pulsar. 
Furthermore, the X-ray spectrum from the Lobe obtained with \xmmnewton\ is 
completely thermal \citep{2004ApJ...617..322S}.

We believed that the bright X-ray emission is the result of 
the interaction between the SNR shock and a molecular cloud complex
and therefore performed an observation with the \chandra\ X-ray Observatory
\citep{2002PASP..114....1W} using the Advanced CCD Imaging Spectrometer
\citep[ACIS,][]{2003SPIE.4851...28G} in order to study this interaction 
region in more detail. 
The first analysis of the high resolution data obtained with ACIS-I in 
combination with new high resolution CO data from the Five College Radio 
Observatory has revealed regions with signs of interaction between the 
SNR shock and CO clouds \citep{2006ApJ...642L.149S}.
In this paper we present the results of the spectral analysis of the entire
northeastern part of the SNR, which was observed with ACIS-I. 

\section{\chandra\ data}

We observed the northeast part of the SNR in an 80~ksec pointing of
\chandra\ with ACIS-I as the prime instrument (ObsID 4626). 
The Lobe and the northeastern part of the remant shell were completely 
covered by the ACIS-I array (Fig.\,\ref{int}). The data were analysed 
with CIAO 4.4.\ and CALDB 4.4.8.

\subsection{Images}\label{ima}

\begin{figure}
\centering
\includegraphics[width=0.5\textwidth]{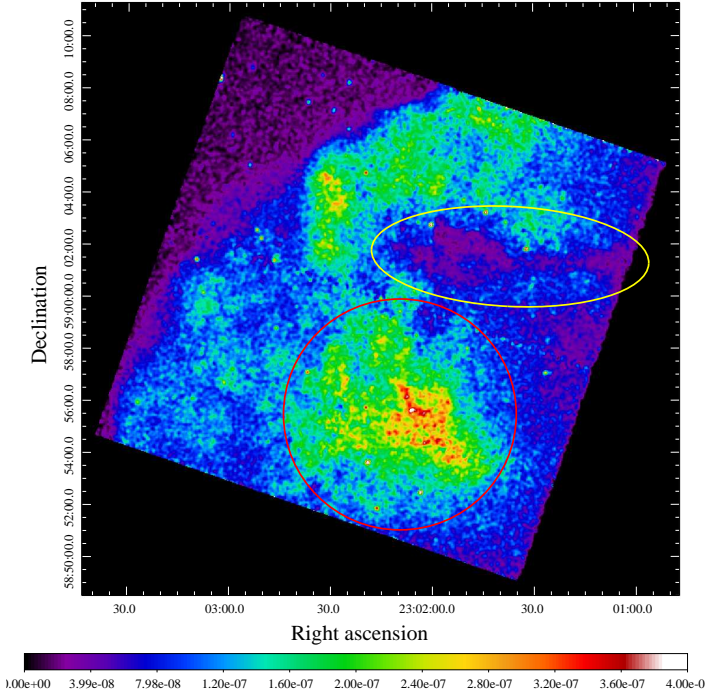}
\caption{\label{int}
Exposure-corrected \chandra\ ACIS-I intensity map (0.35 -- 8.0~keV).
The big red circle indicates the position of the Lobe, the yellow ellipse that 
of the CO arm.
}
\end{figure}

\begin{figure}
\centering
\includegraphics[width=0.5\textwidth]{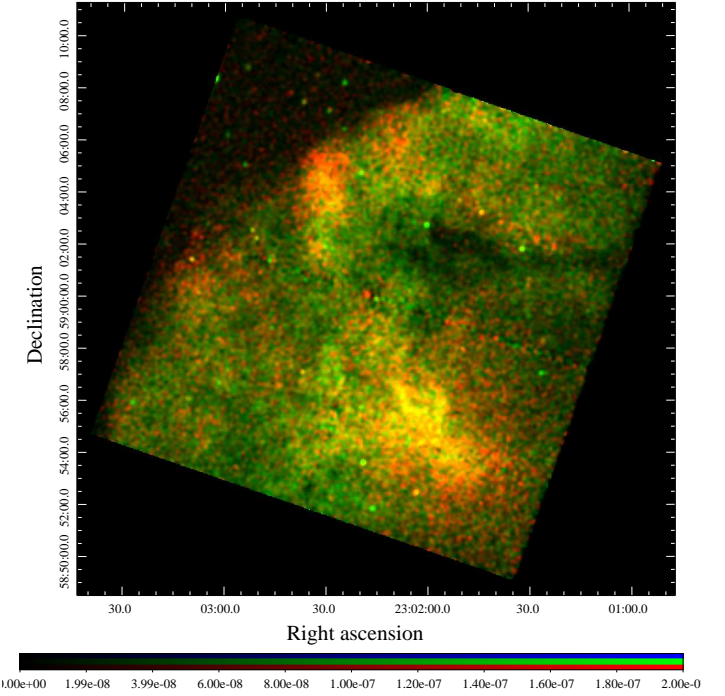}
\caption{\label{rgb}
Two-colour image with red (0.35 -- 0.8~keV) and green (1.0 -- 8.0~keV). 
}
\end{figure}

Figure \ref{int} shows the exposure-corrected intensity map of the ACIS-I 
data (0.35 -- 8.0~keV). The events
were binned with a bin size of 4 pixels and the image has been 
smoothed with a gaussian kernel of 3 and 4 pixels in Figs.\,\ref{int}
and \ref{rgb} (see below), respectively.

Compared to the \xmmnewton\ image \citep{2004ApJ...617..322S}, the \chandra\
image reveals more point sources. Also, structures in the dark region absorbed
by the CO arm, as well as filamentary structures in the Lobe, are resolved. 
The two-colour image (Fig.\,\ref{rgb}), in which the soft band (0.35 -- 0.8~keV)
is presented in red and the hard band (1.0 -- 8.0~keV) in green, shows colour 
variations in the Lobe. 
There is a bright yellow structure in the central part of 
the Lobe, while the outer diffuser parts are green (harder) in the east
and red (softer) in the west. 
The two-colour image shows spectral variations that are either intrinsic or
due to absorption or perhaps a combination of the two.
The diffuse emission in the CO arm appears to have harder spectra
consistent with higher absorption.
A detailed spectral analysis would possibly distinguish between intrinsic spectral
variations and variable absorption.

\begin{figure}
\centering
\includegraphics[width=0.5\textwidth]{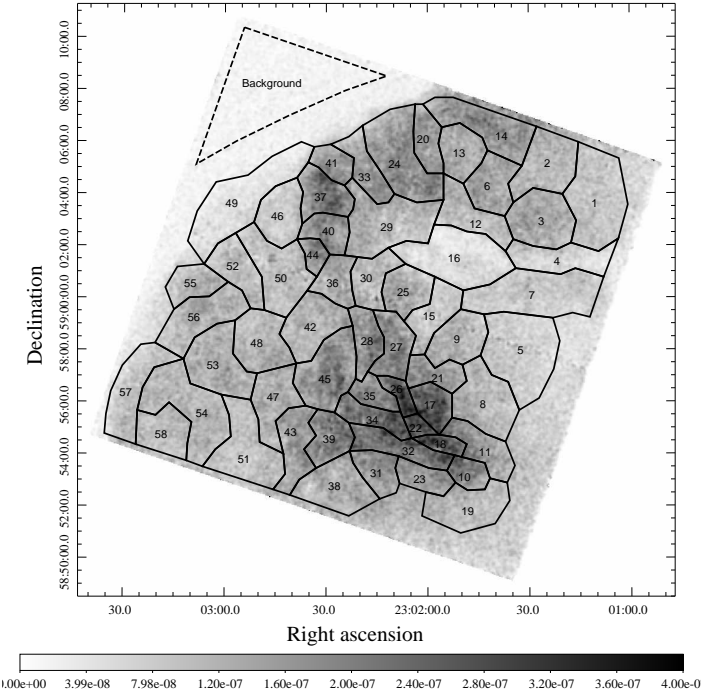}
\caption{\label{regions}
Extracted regions overlaid on the intensity map. Point sources are removed.
}
\end{figure}

For further analysis, we divided the diffuse emission of the SNR into small
regions with similar surface brightness and X-ray colour, and extract spectra
for each region.
To study the spectra of only the diffuse emission, we first performed source
detection on the whole data ({\tt wavdetect}) and excluded all detected point 
and point-like sources in each region.
The regions used are shown in Figure \ref{regions}. The number of counts in
the regions varies from \about4\,000 -- 20\,000 cts.
The spectra are binned with a minimum of 20 counts per bin so that Gaussian
statistics may be assumed in the fitting.

\subsection{Spectral analysis}\label{spec}

We analysed the spectra using the X-ray spectral fitting package XSPEC 
Ver.\ 12.7.1.
The analysis of \xmmnewton\ data showed that the emission from \ctb\ is
thermal with no indication of nonthermal emission \citep{2004ApJ...617..322S}.
Furthermore, the thermal emission suggests that the plasma is not in collisional 
ionisation equilibrium (CIE). Therefore, we fitted the extracted ACIS-I spectra 
with non-equilibrium ionisation (NEI) models. 
The foreground absorbing column density \nh\ was modeled using {\tt TBABS}
\citep{2000ApJ...542..914W}.
We used abundances relative to solar values reported by 
\citet{2000ApJ...542..914W}.

\subsubsection{One-component NEI model}\label{onevnei}

\begin{figure}
\centering
\includegraphics[angle=270,width=0.5\textwidth,clip]{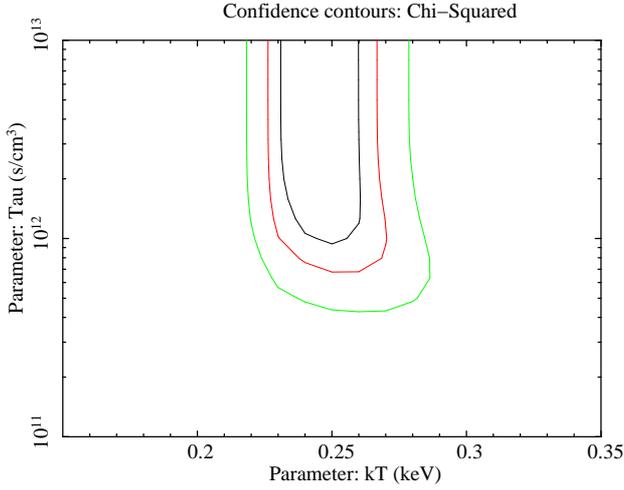}
\caption{\label{cont49}
Contour diagram of the parameters $kT$ and $\tau$ for the fit with one
{\tt VNEI} model of region 49. The contours correspond to confidence
levels of 60\%, 90\%, and 99\%.
}
\end{figure}

First, we fitted all spectra assuming one {\tt VNEI} component, similar to 
what was used to analyse the \xmm\ data \citep{2004ApJ...617..322S}. 
Initially, the abundances were all fixed to solar values.
In regions 49 and 57, which correspond to the outer shock, we get relatively 
good fits with \nh\ = $9.6 (8.8 - 10.5)
\footnote{All errors in this paper are 90\% confidence errors.}
\times\ 10^{21}$~cm$^{-2}$, $1.2 (1.1 - 1.3)$
$\times\ 10^{22}$~cm$^{-2}$, 
$kT$ = 0.25 (0.22 - 0.27)~keV, 0.25 (0.23 - 0.26)~keV, and
$n_{\rm e} t > 7.2 \times\ 10^{11}$~s~cm$^{-3}$,
$> 2.3 \times\ 10^{10}$~s~cm$^{-3}$, 
with reduced $\chi^2 $ = 1.4 for 107 and 99 degrees of freedom for regions 
49 and 57, respectively (see Fig.\,\ref{cont49}).
The best fit values for the ionisation timescale are $n_{\rm e} t = 
4.9 \times\ 10^{13}$~s~cm$^{-3}$ and $2.1 \times\ 10^{13}$~s~cm$^{-3}$
indicating CIE in these regions.

We also freed the abundances, but the one-component description of the
data is not satisfactory in about 25\% of the regions 
with reduced $\chi^2 >$ 1.5, ranging up to reduced $\chi^2 =$ 2.0
(see Fig.\,\ref{spectra}, upper left diagram). 

\subsubsection{Two-component NEI model}

\begin{figure*}
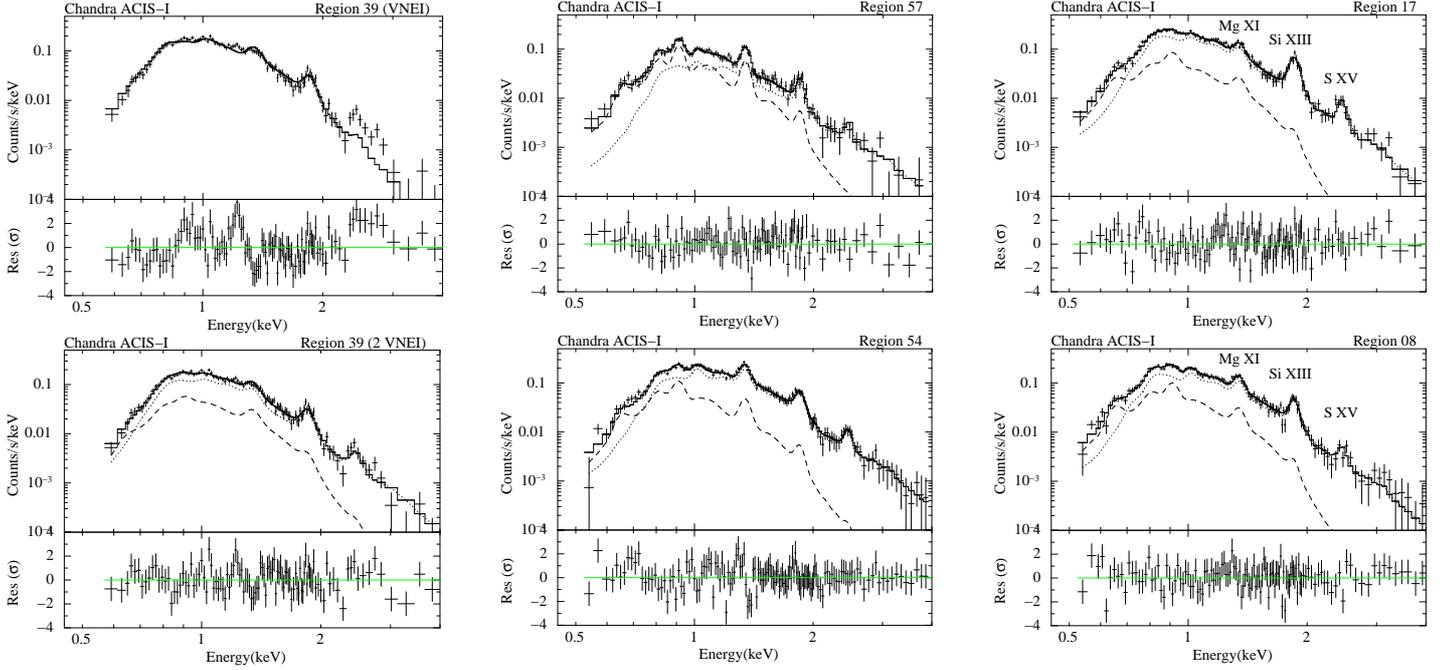

\hspace{-7mm}
\begin{tabular}{ccc}
\includegraphics[angle=270,width=0.33\textwidth]{region_39_vnei_label.ps}  &
\includegraphics[angle=270,width=0.33\textwidth]{region_57_2vnei_label.ps} &
\includegraphics[angle=270,width=0.33\textwidth]{region_17_2vnei_label.ps}
\\[-3mm]
\includegraphics[angle=270,width=0.33\textwidth]{region_39_2vnei_label.ps} &
\includegraphics[angle=270,width=0.33\textwidth]{region_54_2vnei_label.ps} &
\includegraphics[angle=270,width=0.33\textwidth]{region_08_2vnei_label.ps}
\end{tabular}
\caption{\label{spectra}
{\it Left}: 
\chandra\ ACIS-I spectrum of an interior region 39 with the best fit 
single-component {\tt VNEI} model (upper diagram, reduced $\chi^2 =$ 2.0
for 103 degrees of freedom) 
and two-component model (lower diagram, reduced $\chi^2 =$ 1.1
for 100 degrees of freedom).  
{\it Middle}:
Spectrum of the outermost region 57 in the east (upper) 
and an interior region 54 (lower) with the best fit 2 {\tt VNEI} model.  
The relative flux of the ejecta component (dotted) with respect to the ISM 
component (dashed) in region 57 is lower than in most other regions. 
{\it Right}:
Spectrum of region 17 with enhanced \sixiii\ emission 
(upper diagram). The \mgxi, \sixiii, and \sxv\ triplets are marked. For comparison, 
the spectrum of region 08, which is the brightest region in the Lobe next to 
region 17, is shown (lower diagram). 
The best fit 2 {\tt VNEI} model is additionally plotted in the diagrams.
}
\end{figure*}

Therefore, in the next step we assumed two thermal {\tt VNEI} components: 
1) to model the emission of the shocked ISM
and 2) to verify the existence of emission from shocked ejecta.
This two-component model improves the fits in a number of regions
(see below), especially in region 39, which had a reduced $\chi^2 =$ 2.0
(103 degrees of freedom) for the one-component {\tt VNEI} 
model and can be fitted with a reduced $\chi^2 =$ 1.1
(100 degrees of freedom)
with a two-component {\tt VNEI} model (Fig.\,\ref{spectra}, left panels).

For the ISM component, we use $kT_1 = 0.25$~keV and $n_{\rm e} t_1 =
1 \times 10^{12}$~s~cm$^{-3}$ as starting values, representing the best fit 
parameters of the single-component {\tt VNEI} model for regions 49 and 57.
All abundances of this component were fixed to solar values.
The second component, which was introduced to model the ejecta emission,
was fitted with variable abundances for elements showing strong emission lines. 
The fits required a higher temperature of $kT_2 \approx\ 0.6$~keV. 
The ejecta component with the higher temperature
dominates the spectrum for energies higher than $\sim$1~keV in most of the 
spectra (see Fig.\,\ref{spectra}, middle panels).
The abundances of Mg, Si, S, for which emission lines are visible in the 
spectra, as well as Fe, were freed for the second (ejecta) model component and 
fitted. In some regions, higher abundances are found for Si and 
Fe in particular (see Fig.\,\ref{spectra}, right panels). 

\subsubsection{F-test}\label{ftest}

\begin{figure*}
\hspace{-3mm}
\begin{tabular}{cccc}
\includegraphics[width=0.32\textwidth]{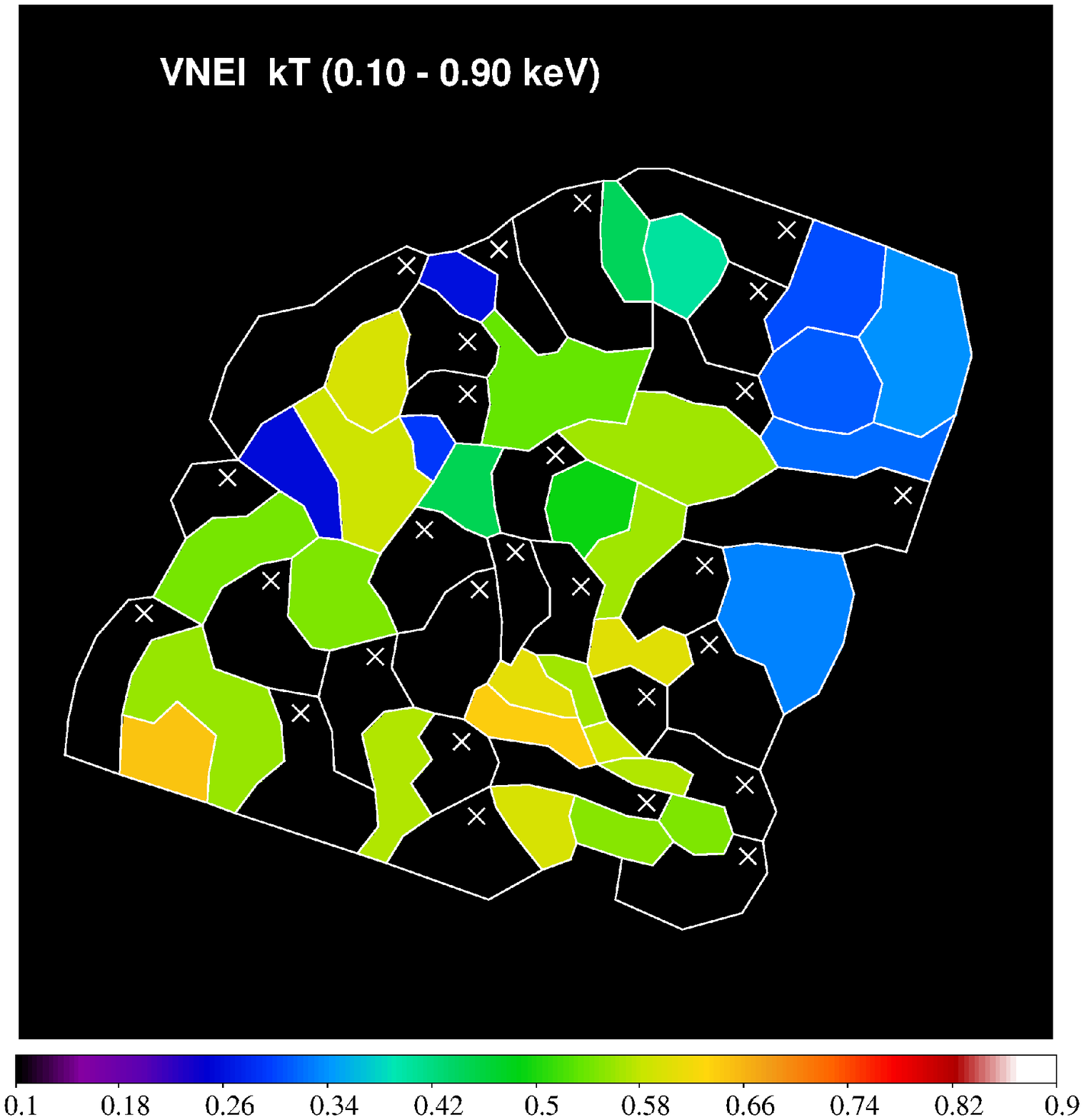}& &
\includegraphics[width=0.32\textwidth]{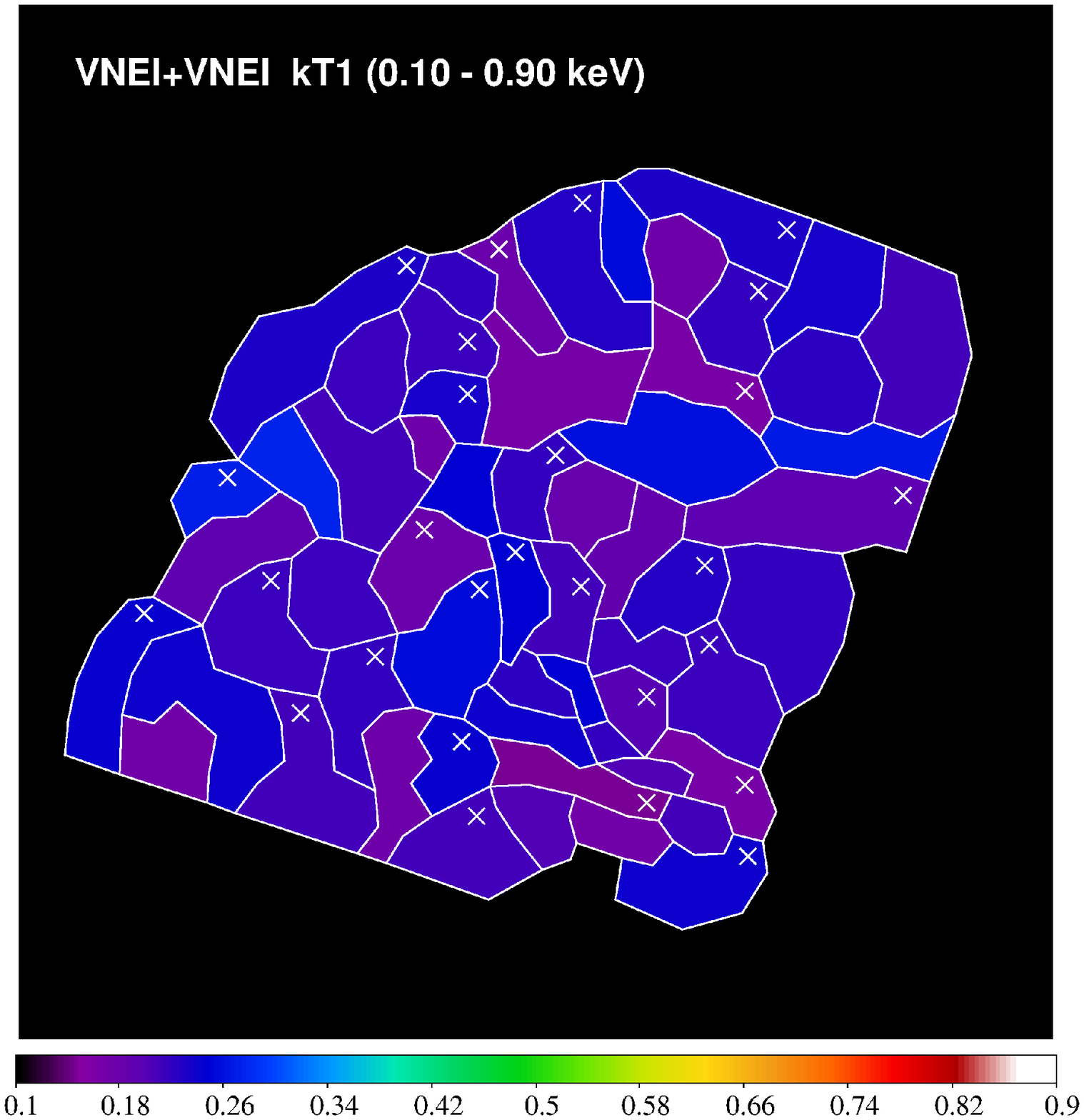}&\hspace{-5mm}
\includegraphics[width=0.32\textwidth]{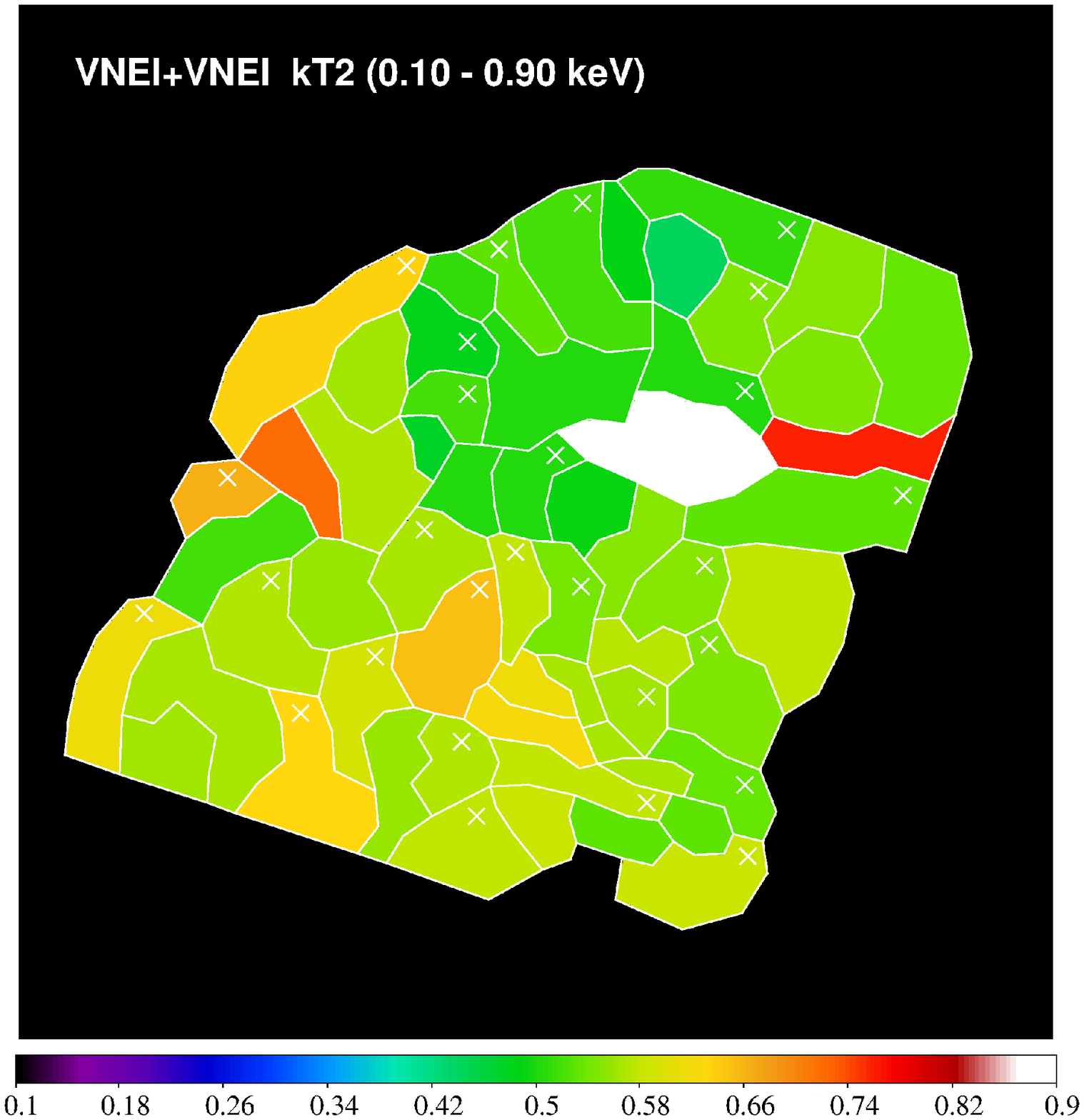}\\[3mm]
\includegraphics[width=0.32\textwidth]{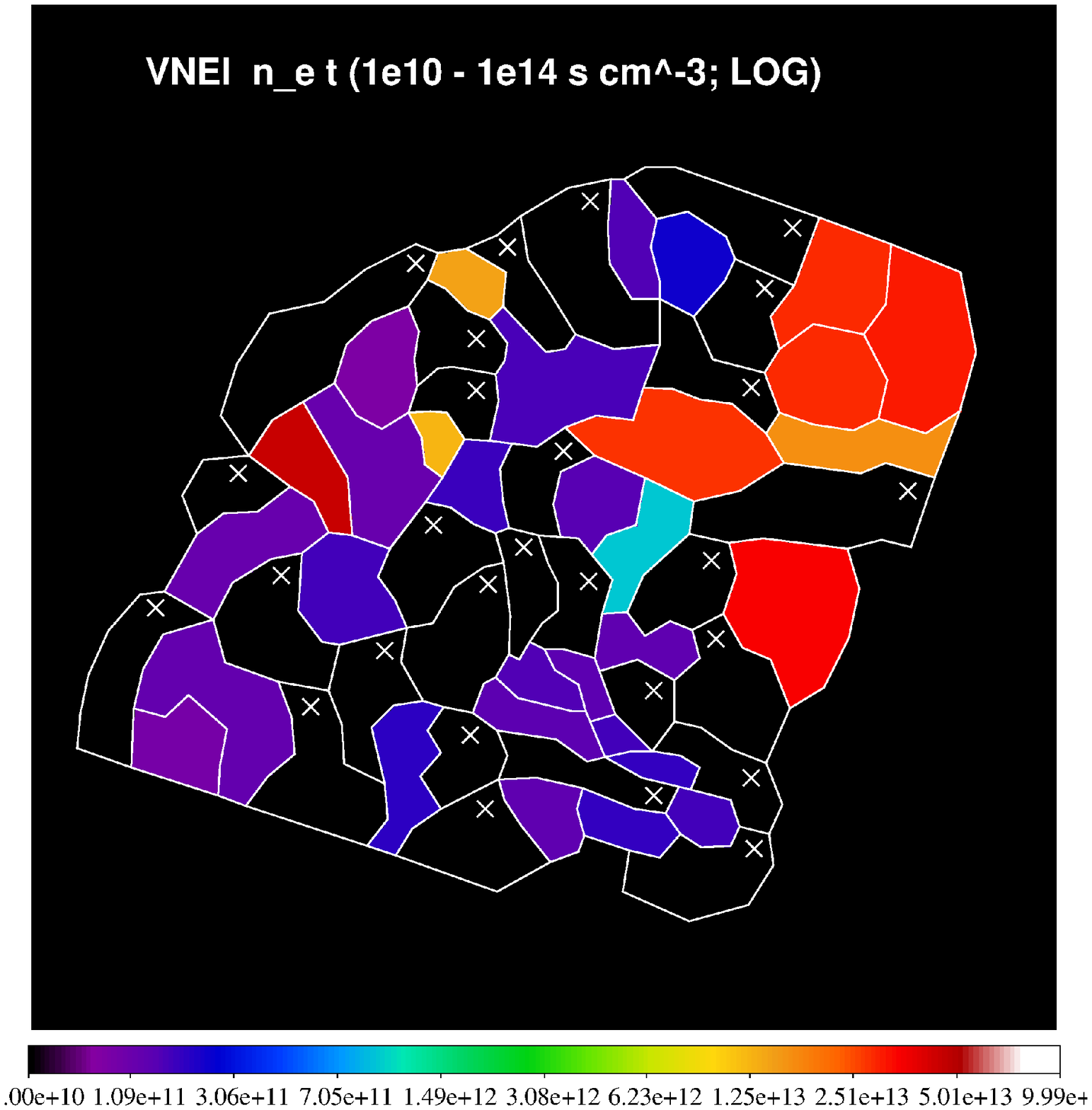}& &
\includegraphics[width=0.32\textwidth]{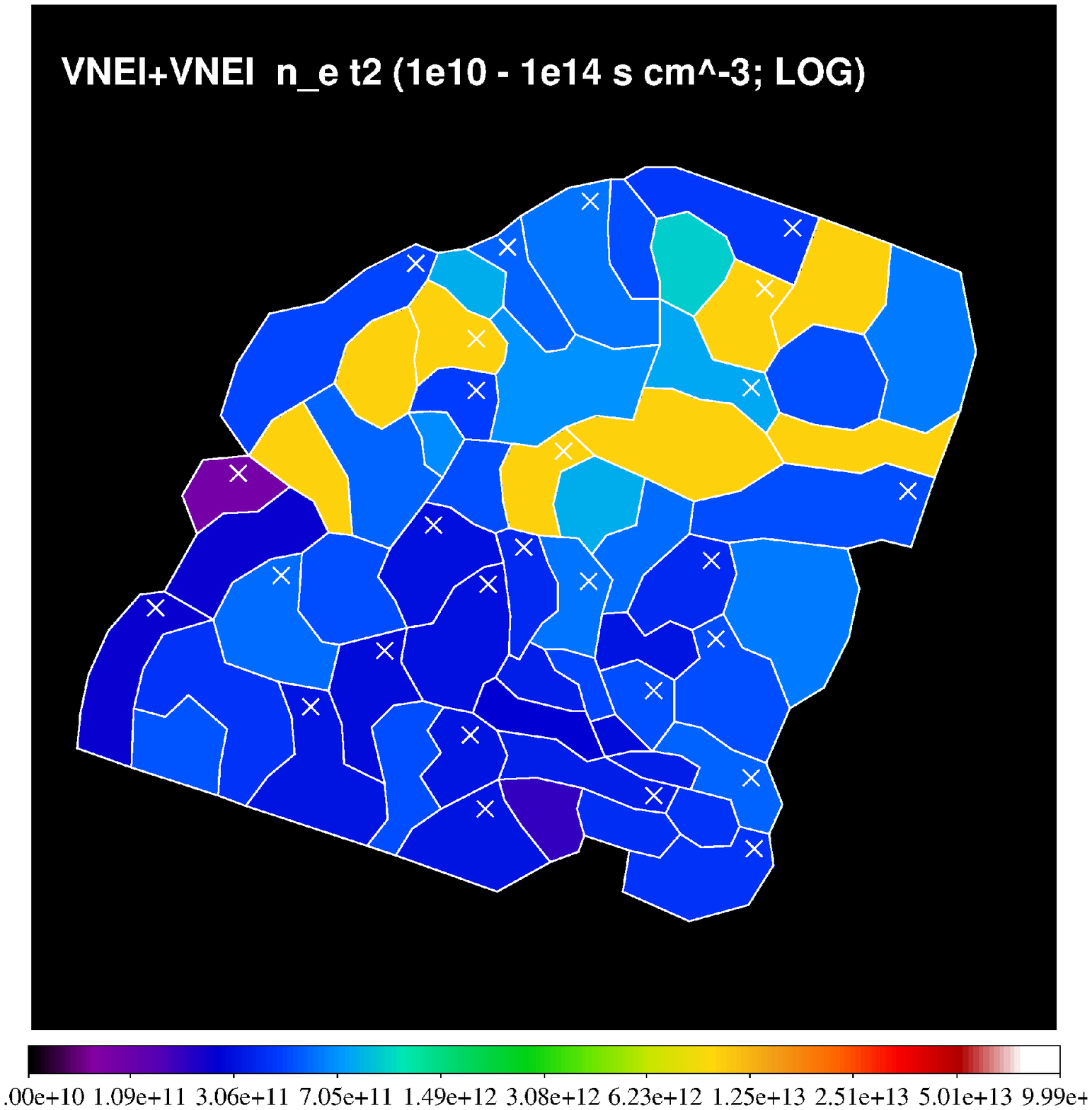} & \\[3mm]
\includegraphics[width=0.32\textwidth]{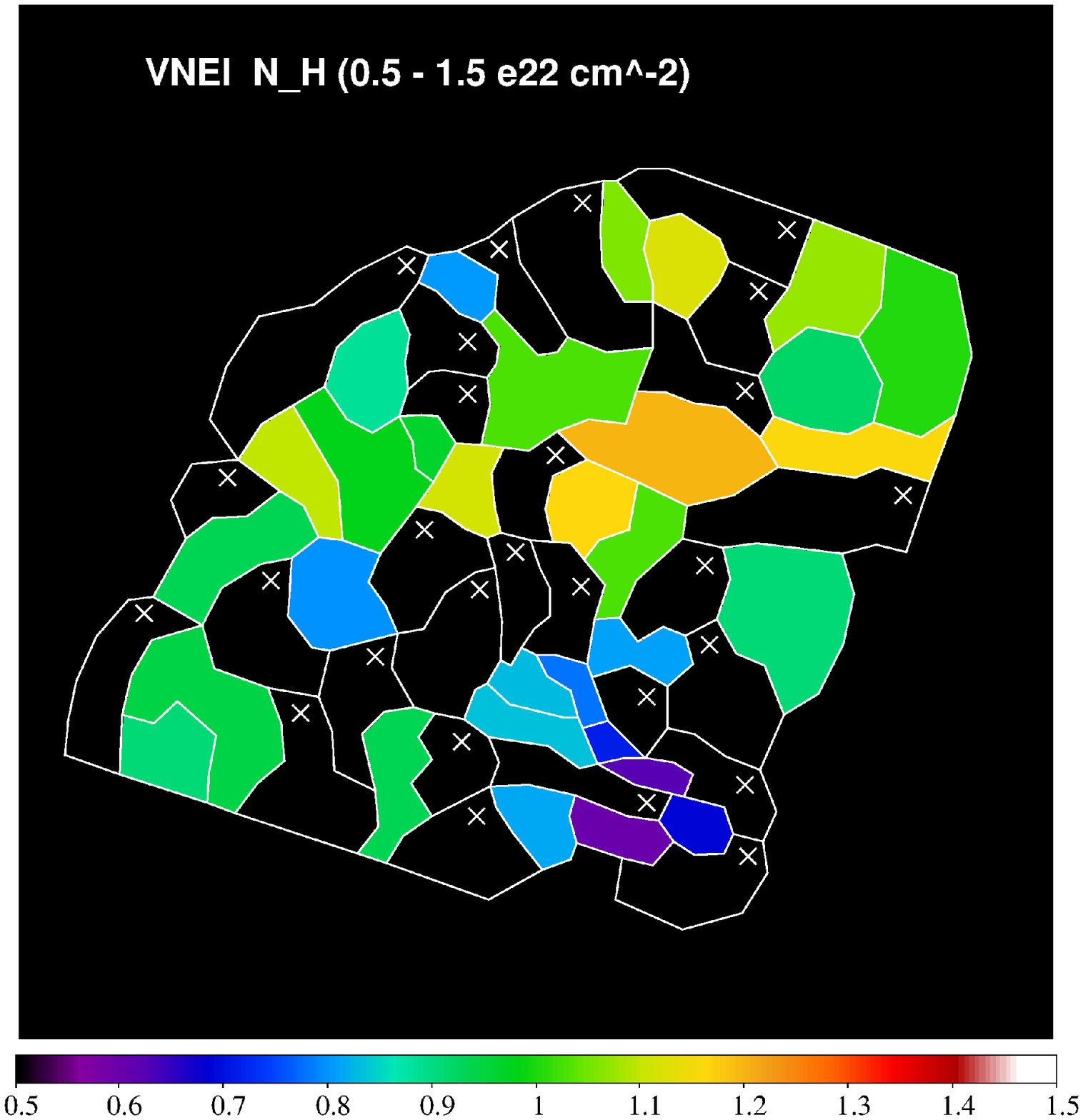}& &
\includegraphics[width=0.32\textwidth]{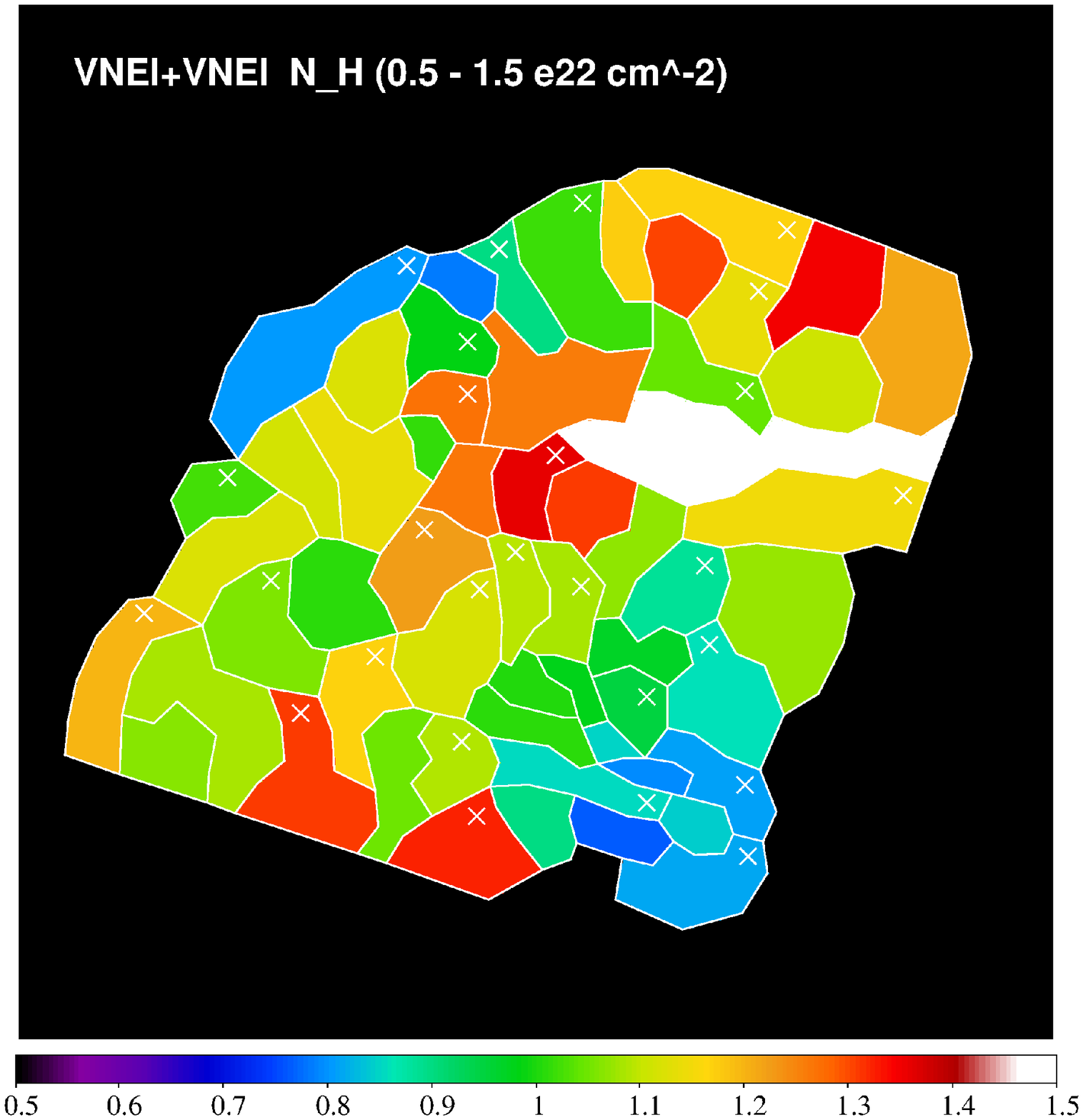} & 
\end{tabular}
\caption{\label{ftestpar}
{\it Top:} Best fit temperature parameters for the one-component 
{\tt VNEI} model ($kT$) in regions in which the F-test indicated that the
one-component model fits the spectrum sufficiently well (left) and
temperature parameters $kT_1$ and $kT_2$ for the two-component
fits for all regions (middle and right, respectively).
{\it Middle:} Best fit values for the ionisation timescale $\tau$ for the 
one-component {\tt VNEI} model (left) and the 
ionisation timescale $\tau_2$ for the dominating ejecta component
for the two-component fits (right).
{\it Bottom:} Best fit values for the foreground \nh\ for the one-component 
{\tt VNEI} model (left) and for the two-component fits (right). 
The crosses mark the regions in which the F-test indicated that the
two-{\tt VNEI} component model yields a better fit (see Sect.\,\ref{ftest}).
}
\end{figure*}

To verify how much the fits improve by including a second {\tt VNEI} 
component, we performed an F-test for the one-component {\tt VNEI} and 
two-component {\tt VNEI} models for all regions. In 27 out of 58 
regions, the two models differ with probabilities higher than 99.9\%. 
In order to visualize the variations of the spectral properties, we created a
map of the spectral parameters by filling each region with the corresponding
parameter value.
In the top row of Fig.\,\ref{ftestpar} we show the distribution of 
the temperature parameter $kT$ of the one-component {\tt VNEI} model for 
regions in which the F-test indicated no necessity of an additional component 
(left) and the distributions of $kT_1$ (ISM) and $kT_2$ (ejecta) of the 
two-component {\tt VNEI} model (middle and right, respectively). 
Regions in which the F-test confirmed an improvement of the fits after
including the second {\tt VNEI} component are marked with white crosses.
The middle and bottom rows in Fig.\,\ref{ftestpar}
show the same for the ionisation timescales $\tau$ (left) and $\tau_2$ (right) 
and the foreground column densities \nh\ for the one-component (left) and 
two-component (right) fits.

\section{Discussion}\label{discuss}

\begin{figure}
\centering
\includegraphics[width=0.41\textwidth,bb=43 0 446 355,clip=]{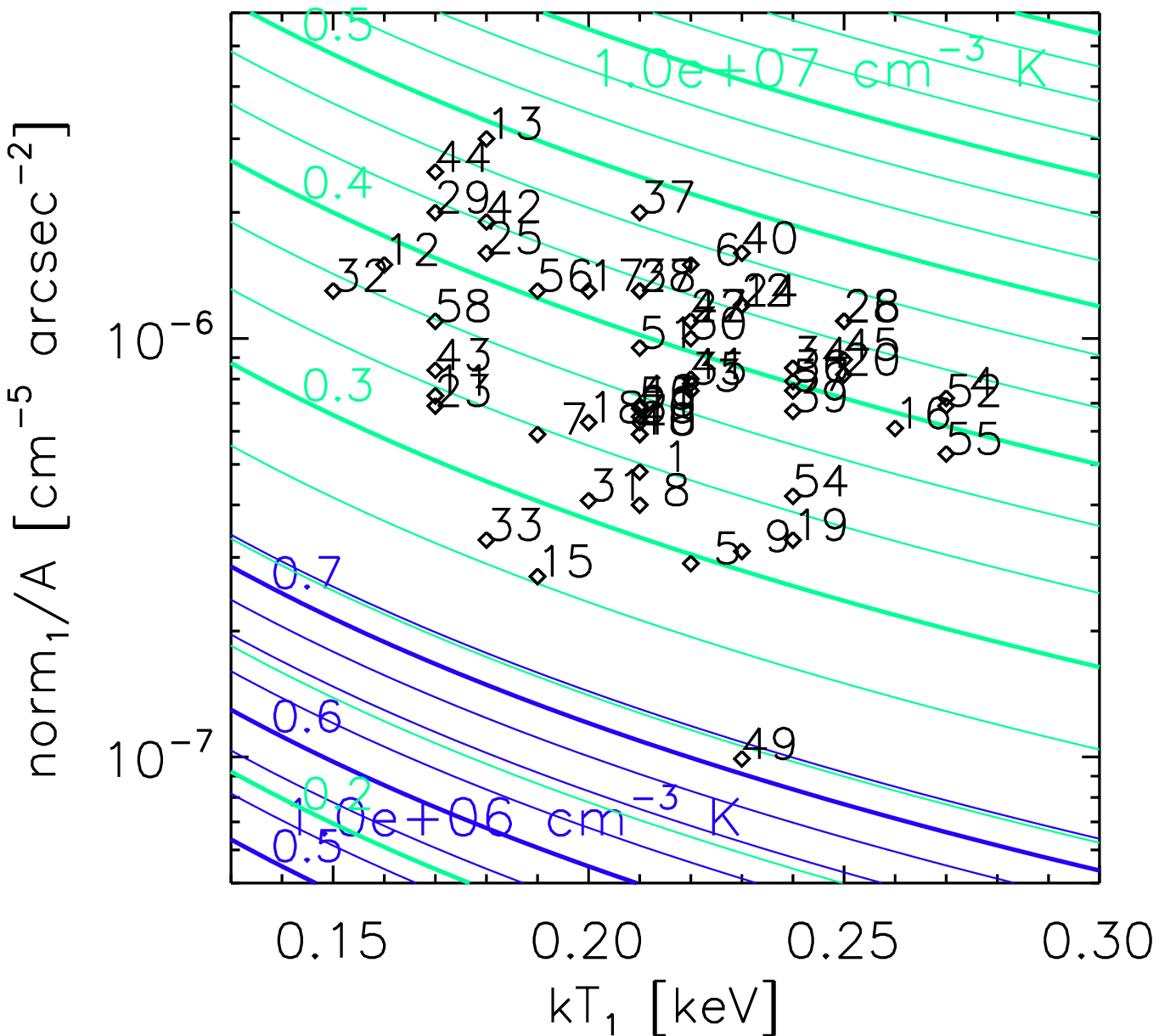}
\includegraphics[width=0.41\textwidth,bb=43 0 446 355,clip=]{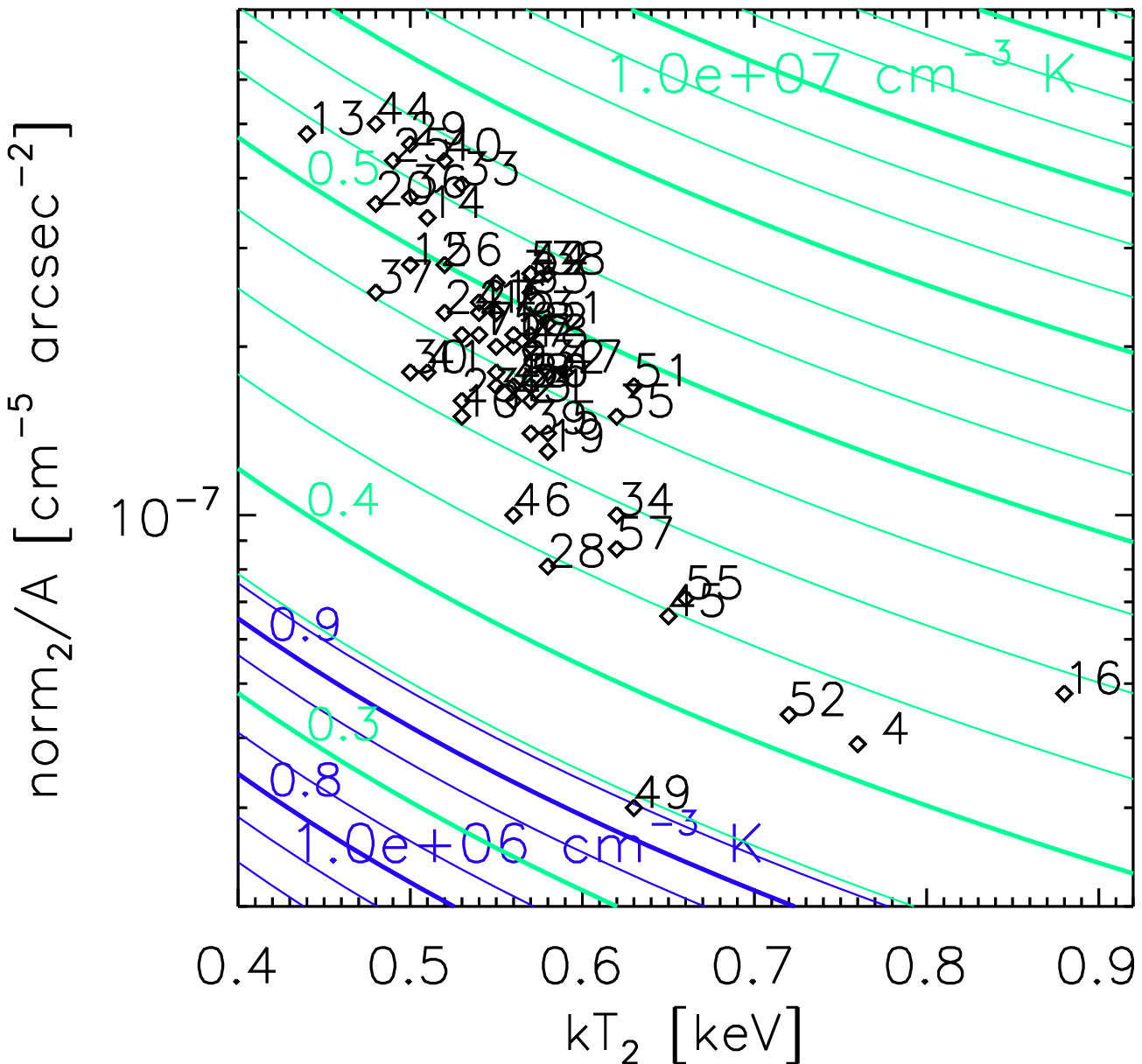}
\includegraphics[width=0.41\textwidth,bb=43 0 446 355,clip=]{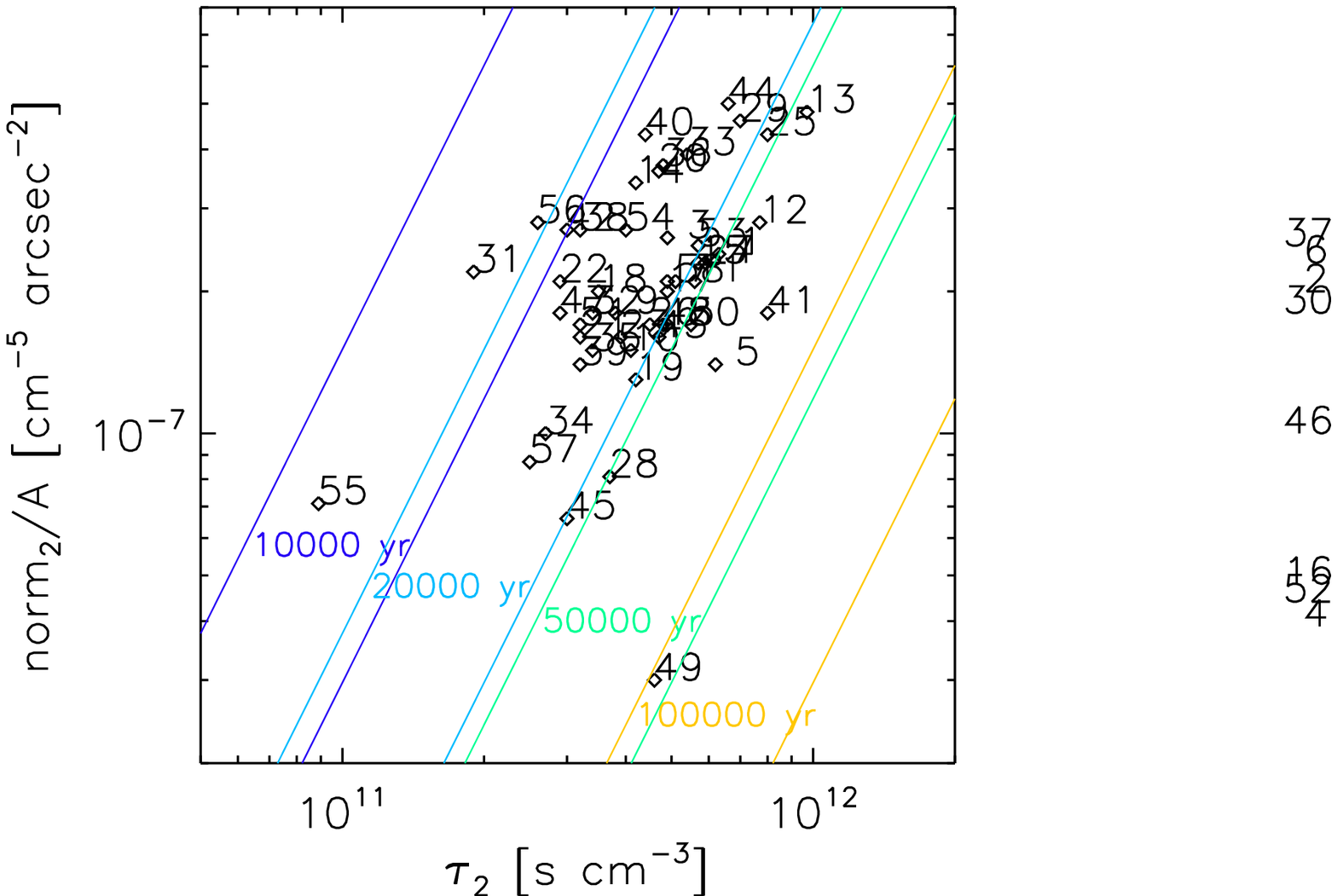}
\caption{\label{scatter}
Distribution of the parameter values for the fit with two {\tt VNEI} components.
The labels of the data points are region numbers.
The lines in the diagrams for surface brightness $norm_{1,2}/A$ against
temperature $kT_{1,2}$ (upper and middle panels) show the isobars for 
$p/k = 10^6$~cm$^{-3}$~K (blue) and $10^7$~cm$^{-3}$~K (green) for different 
values of the filling factors in the remnant and the size of the emitting 
volume along the line of sight.  
The line for each filling factor with the lowest $norm_{1,2}/A$ are marked 
with a thicker line and labelled with the $f_{1,2}$ value.
The lines in the diagram showing the surface brightness as $norm_2/A$ against the 
ionisation timescale $\tau_2$ (bottom panel) are isochrones for 
$t =$ 10\,000, 20\,000, 50\,000, and 100\,000~yr. 
A range for all isochrones is shown for each $t$ covering the range of 
variation in the depth of the emitting volume.}
\end{figure}

The analysis of the \chandra\ ACIS spectra of the northeastern part of the SNR
\ctb, divided into small regions, has shown that the spectra can be well fitted
by a one-component {\tt VNEI} model in many regions. However, in 50\% of the 
regions a two-component model consisting of two thermal {\tt VNEI} models with 
two different temperatures improves the fit significantly. In this model, the 
first component is used to describe the emission from the shocked ISM, while 
the second component reproduces the emission from the ejecta. 
Therefore, we believe that there are at least two emission components in
all regions. In cases in which the one-component model yields a sufficiently 
good fit, the temperature and the ionisation timescale most likely tend toward 
values of the component of the multi-temperature spectrum that dominates the 
emission.
We base our discussion on the results obtained with the two-component
spectral fits for all regions.
The diagrams in Fig.\,\ref{scatter} show the relations between some parameters 
derived from the fits (see following subsections for further discussion). 

The temperature of the ISM component (Fig.\,\ref{ftestpar}, upper middle) is 
comparable in all regions and is low ($\sim0.1 - 0.3$~keV), while the temperature 
of the ejecta component is higher ($\sim0.4 - 0.9$~keV, Fig.\,\ref{ftestpar}, upper right).
The ionisation timescale of the ejecta component is $\tau_2 = 10^{11-12}$~s\,cm$^{-3}$
(Fig.\,\ref{ftestpar}, middle panels) except for regions at and around the CO arm, in which
$\tau_2$ tends to be higher.
The higher values for $\tau_2$ in the CO arm regions indicate a higher density 
as we can assume that these regions have been shocked at a similar time to the 
rest of the remnant.

The \nh\ images (Fig.\,\ref{ftestpar}, lower panels) show that the foreground 
absorption is 
high near the CO arm (\nh $> 1.1 \times 10^{22}$~cm$^{-2}$).
Around and, in particular, west of the Lobe, the foreground absorption is lower
(\nh $< 1.0 \times 10^{22}$~cm$^{-2}$), while it seems to be higher southeast
of the Lobe.

\subsection{Filling factors and ejecta mass}

\ctb\ is an evolved SNR in the Sedov phase. In such a remnant, which
also shows interactions with denser molecular clouds, we can assume that
the reverse shock has already propagated to the centre 
and thus heated most or all of the ejecta. 
The spectral model provides us with the normalisation parameter
\begin{equation}\label{eq_norm}
norm_{1,2} = \frac{10^{-14}}{4 \pi D^{2}} \times \int n_{1,2}~n_{e~1,2}~f_{1,2}~dV,
\end{equation}
with $f_1$ and $f_2$ being the filling factors of the ISM and ejecta 
components, respectively ($f_1 + f_2 = 1$).
For the distance to the SNR, we use $D = 3.0\pm0.5$~kpc as estimated by 
\citet{2002ApJ...576..169K} and confirmed by Kothes \& Foster 
(\citeyear{2012ApJ...746L...4K}, $3.2\pm0.2$~kpc).
To calculate the filling factor for the ejecta component $f_2$
we assume pressure equilibrium between the two thermal components.
Applying momentum conservation for the shock propagating through these
components, we get
\begin{equation}\label{eq_f2}
f_2 = \left(\frac{norm_1~T_1^2}{norm_2~T_2^2}+1\right)^{-1}
\end{equation}
\citep[see][]{1999A&A...342..839B}.

\begin{figure}
\centering
\includegraphics[width=0.32\textwidth]{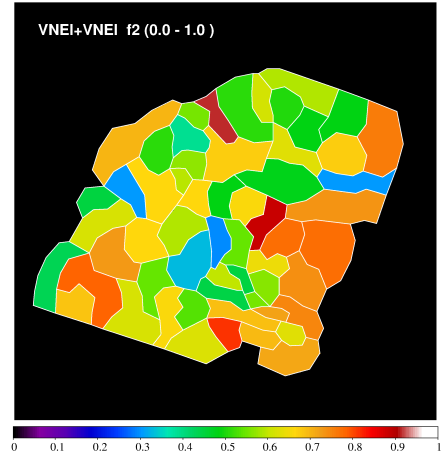}
\caption{\label{mapf2}
Distribution of the filling factor for the ejecta component $f_2$
derived from the fits with the two-component {\tt VNEI} model.
}
\end{figure}

The resulting parameter map for the filling factor of the ejecta
component of the two-component {\tt VNEI} fits
is shown in Fig.\,\ref{mapf2} for all regions.
We indeed see a trend in the parameter value distribution map in
Fig.\,\ref{mapf2}, in which the inner (western) regions tend to have higher
$f_2$ values than the outer regions.

We can estimate the total ejected mass using 
the filling factor $f_2$ and the density $n_2$ derived for the ejecta component
from Eq.\,\ref{eq_norm}.
We assume that the SNR is a half sphere with a radius of 
$18\farcm5\pm1\farcm0$ as determined from the \xmm\ EPIC images by 
\citet{2004ApJ...617..322S}. For D = 3.0 $\pm$ 0.5~kpc,
$R 
= 16\pm3$~pc
= $5.0\pm0.8 \times 10^{19}$~cm.
Since we have performed spectral analysis using a model with two thermal 
components only in regions that have been observed with \chandra\ ACIS-I,
we use the median values of $f_2$ and $n_2$ from the two-component fits to 
estimate the total mass of the ejecta. With $f_{\rm 2, median} = 0.44$ and
$n_{\rm 2, median} = 0.62$~cm$^{-2}$, the total ejecta mass under the 
assumption of a half sphere as seen, e.g., in the \xmm\ mosaic image, is
$M_{\rm ejecta} = 60~M_{\sun}$.
The density $n_{\rm 2}$ in the regions observed with \chandra\ varies
from 0.23 to 0.81~cm$^{-2}$, while the filling factor $f_2$ has values 
between 0.29 and 0.91. Therefore, these estimates have uncertainties of
$\sim$50\%, whereas the volume may also have an uncertainty of the same
order
since the X-ray emitting hot gas might not fill the entire volume
applied for the estimate. In addition the ejecta might be clumped and
thus not uniformly distributed in the assumed volume. 
If we suppose that the ejecta only fills 30\% of the volume, the ejecta mass 
will be as low as $M_{\rm ejecta} = 20 M_{\sun}$.
Therefore, one should keep in mind that $M_{\rm ejecta}$ is a very crude 
estimate. 
However, even the lower limit of $M_{\rm ejecta} = 20 M_{\sun}$ would
rule out a Type Ia supernova as the origin of \ctb.


\subsection{Pressure}

In the $norm_{1,2}/A$-$kT_{1,2}$ diagrams in Fig.\,\ref{scatter} 
(upper two diagrams) we have plotted
lines of constant pressure with $p/k = 10^6$~cm$^{-3}$~K (blue) and 
$10^7$~cm$^{-3}$~K (green) for different values of filling factors and depths 
of the emitting volume in the SNR. These lines were calculated from the best 
fit parameters assuming pressure equilibrium 
between the ISM and ejecta components. All the data points are consistent with
a thermal pressure of $p/k \approx 10^7$~cm$^{-3}$~K, corresponding to 1.4 
$\times 10^{-13}$~dyne~cm$^{-2}$. The scatter is mainly due to the different 
filling factors and the uncertainties in the size of the emitting volume along 
the line of sight.

%

\subsection{Age estimate}

In the $\tau_2$-$norm_2/A$ plot in Fig.\,\ref{scatter} (lower diagram), 
we plot the ionisation timescale $\tau_2$ of the ejecta component versus the 
normalisation $norm_2$ per area $A$ of each region. 

To study the dependence of the X-ray surface brightness, measured as
the normalisation parameter per area $norm_2/A$, and the ionisation 
timescale $\tau_2$, we can replace $n_{\rm e}$ by $\tau/t$ in 
Eq.\,\ref{eq_norm}. We then get $norm_2/A$ as a function of $\tau$
for a given $t$.
The value of $norm_2$ depends on the emitting volume, thus on
the depth of the volume along the line of sight for each region.
Assuming that the AXP is located at the centre of the circle corresponding
to the SNR shell, we calculated the minimum and maximum distances from
the AXP for each region.
The calculated isochrones for different $t$ values are also shown
in the $\tau_2$-$norm_2/A$ plot. The two lines for each time $t$
show the possible ranges
for the depth of the remnant. This diagram indicates that the 
age of the shocked plasma, derived from the ejecta component (2) is 
$\sim2 \times 10^4$~yr. 

The fits have shown that the temperature of the ISM component (1) is 
between 0.1~keV and 0.3~keV. For the outer regions 49 and 57, the 
single component {\tt VNEI} fit yields $kT$ = 0.25~keV (Sect.\,\ref{onevnei}).
The high and not well-constrained values of the ionisation timescale
$\tau$ in the single-component fits, as well as the $\tau_1$ in the
two-component fits, suggest that the plasma of the shocked ISM is close 
to CIE. If we assume that this temperature is more or less representative 
of the shocked plasma, the blast wave velocity is
\begin{equation}\label{eq_vel}
v = \sqrt{\frac{16 kT_1}{3 \bar{m}}},
\end{equation}
with a mean mass per free particle for a fully ionised plasma of 
$\bar{m} = 0.61~m_{\rm p}$. 
%
%
For a temperature of $kT_1 = 0.25\pm0.03$~keV, we get
$v = 460\pm30$~km~s$^{-1}$.
The radius is $R = 5.0\pm0.8 \times 10^{19}$~cm \citep{2004ApJ...617..322S}.
Using the Sedov similarity solution
\citep{1959book..........S,1950PRSLA..201..159T,1947LASLTS.......vN}, 
the age of the remnant can be estimated as
\begin{equation}\label{eq_t}
t = \frac{2R}{5v}.
\end{equation}
%
%
We thus obtain $t = (14\pm2) \times 10^3$~yr from the ISM component,
which is a little higher than the value obtained from the \xmm\ data 
\citep{2004ApJ...617..322S} for which we only assumed one spectral 
component. Interestingly, this new value for the age of the SNR is in 
agreement with the age estimate obtained from the fits of the ejecta
component.

\subsection{Ejecta}\label{ejecta}

\begin{figure*}
\centering
\includegraphics[width=\textwidth]{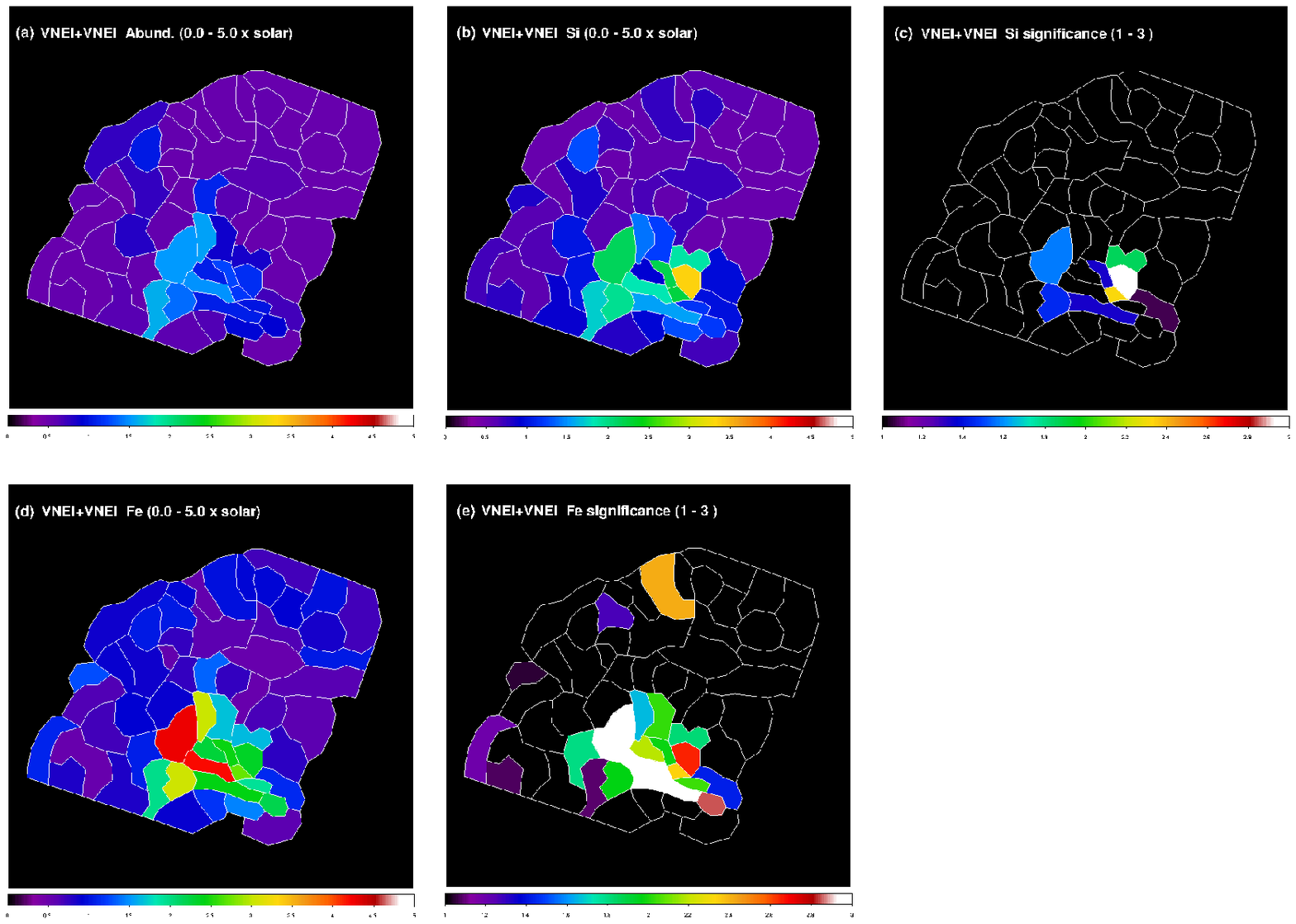}
\caption{\label{mapabund}
Best fit abundances assuming a thermal model consisting of two {\tt VNEI} components.
The shown parameters are: abundances for elements other than Si, S, and Fe fitted 
for the ejecta emission (a), Si abundance fitted for the ejecta emission (b) and
its significance calculated as (abundance of the Si - abundance of the other 
elements)/error of the abundance of Si (c), Fe abundance and its significance
(d and e, respectively).}
\end{figure*}

As can be seen in the Si-abundance and Fe-abundance distribution images 
in Fig.\,\ref{mapabund}, the abundances measured for the ejecta emission are 
all comparable to or lower than solar values, except for the Lobe. While the 
Ne, Mg, and S abundances seem to agree with those of the other elements, 
Si and Fe abundances are higher, especially in and around the Lobe.
The enhanced abundances in the Lobe suggest that its emission has a 
contribution from an ejecta clump or a conglomeration of ejecta clumps and 
shocked clouds.
There is one particular region (region 17) with significantly enhanced 
Si abundance within the Lobe (3.3 [2.6 -- 4.0], see Fig.\,\ref{mapabund}b, c). 
The spectrum of region 17 is shown together with that of an adjacent region 
with lower Si abundance in Fig.\,\ref{spectra} (right panels).

Another possible process that can cause enhanced emission for particular
elements in thermal plasma is charge exchange. However, to produce 
charge exchange emission there must be a phase of cold neutral gas 
next to or inside the hot X-ray emitting plasma. CO data show no emission at 
the position of the Lobe or west of it, where the Si abundance enhancement 
is observed in X-rays \citep{2006ApJ...642L.149S}. For a mature SNR like 
\ctb\ we can assume that the reverse shock has propagated all the way through
the SNR and, therefore, there should be no unionised gas in the interior of 
\ctb. Furtheremore, if the enhanced emission of Si is due to charge exchange,
one would expect even higher indication of charge exchange for O or Ne for 
solar or cosmic abundances. Since no emission enhancement is observed for 
these lower Z elements, we can rule out that the
enhanced emission of particular elements is caused by charge exchange.

The detection of ejecta emission is also very important for the study of
the AXP 1E\,2259+586, as it will give information on its progenitor.
Together with soft gamma-ray repeaters (SGRs) and the sub-class of
rotation-powered high-magnetic field pulsars, the AXPs are believed to
form a class of neutron stars with extremely high magnetic fields. 
So far, there are only a few confirmed associations between SNRs and
high-magnetic field neutron 
stars\footnote{\it http://www.physics.umanitoba.ca/snr/SNRcat/}: 
SNR G292.2--0.5 and the pulsar J1119--6127,
SNR Kes 75 and the pulsar J1846--0258, 
SNR Kes 73 and AXP 1E\,1841--045, 
SNR CTB 109 and AXP 1E\,2259+586, 
SNR G327.2--0.1 and AXP 1E\,1547.0--5408,
SNR G337.0--0.1 and SGR 1627--41,
SNR G042.8+00.6 and SGR 1900+14, and
SNR N\,49 and SGR 0526--66 in the LMC.
Recently, \citet{2012ApJ...748..117P} detected ejecta emission in a 
deep observation of the SNR N\,49 with \chandra\ (120~ks exposure).
However, the element abundance ratio derived from the emission seems
to be more consistent with a Type Ia supernova, which would rule out
the association between the SNR and the SGR, and requires further 
investigation. 
\citet{2011ApJ...732..114L} measured enhanced abundances of Mg, Si, and
S in SNR Kes 73 associated with the AXP 1E~1841--045,
whereas \citet{2012arXiv1210.5261S} reported on possible abundance 
enhancement in the X-ray spectra of SNR G292.2--0.5 and SNR Kes 73.

Ejecta emission is also detected in other middle-aged remnants, e.g. in the 
Galactic SNR G349.7+0.2 \citep{2005ApJ...618..733L}, which is interacting with 
a molecular cloud,
or SNRs in the Large Magellanic Cloud (LMC) 
0548--70.4, 0534--69.9 \citep{2003ApJ...593..370H}, and 
N49B \citep{2003ApJ...592L..41P}.
Clumping of ejecta seems to be common in core-collapse SNRs. Ejecta 
clumps have also been found in evolved SNRs like the Vela, both in the interior
\citep{2008ApJ...676.1064M} and outside the main shell 
\citep{1995Natur.373..587A}.
\citet{2003A&A...408..621K} have shown
in two-dimensional simulations that macroscopic mixing occurs in core-collapse
supernova explosions forming clumps of metals. Rayleigh-Taylor 
instabilities grow in the layers of metals during the expansion, and these 
layers fragmentate into ejecta clumps.

\begin{figure}
\centering
\includegraphics[width=0.5\textwidth]{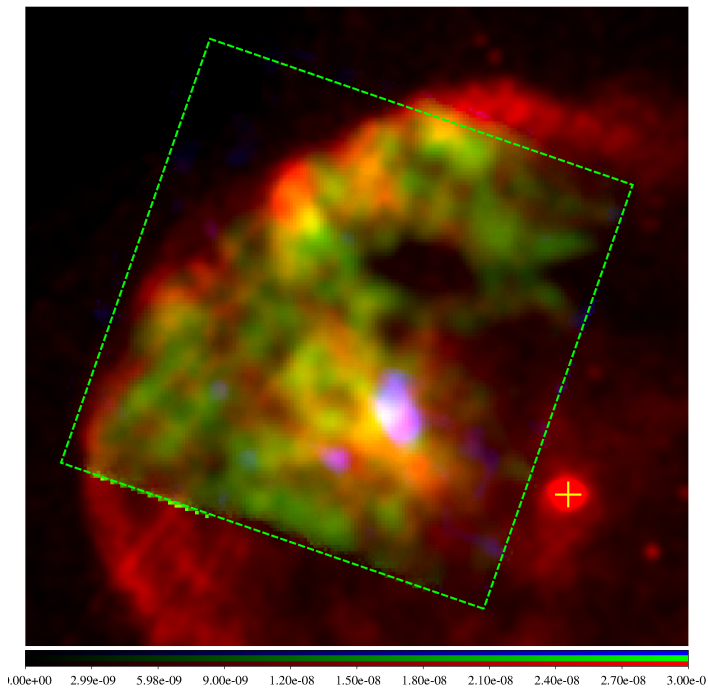}
\caption{\label{softMgSi}
Three colour image consisting of red image for soft band \xmm\ image
\citep[0.3 -- 0.9~keV,][]{2004ApJ...617..322S}, 
green image for Mg (1.25 -- 1.45~keV) -- continuum, and blue image for
Si (1.7 -- 2.1~keV) -- continuum. The Mg and Si images were created from
the \chandra\ data. The green dashed box indicates the field of view of 
the \chandra\ ACIS-I. 
The yellow cross indicates the present position of the AXP
1E\,2259+586.
}
\end{figure}

We created narrow band images for the elements Mg and Si, as prominent 
emission lines of these elements are visible in the spectra of some regions
(Sect.\,\ref{spec}). The continuum emission has been subtracted from the
line emission. 
Figure \ref{softMgSi} shows a three-colour image with 
red for the soft band \xmm\ image (0.3 -- 0.9~keV) presented by 
\citet{2004ApJ...617..322S}, 
green for Mg emission 
([1.25 -- 1.45~keV band] -- surrounding continuum), and blue for
Si emission ([1.7 -- 2.1~keV band] -- continuum). 
There is one small region bright in Si
emission, covered by the extraction region 17 with enhanced
Si line emission (see Sect.\,\ref{spec}). 
The Mg emission is distributed broadly; soft emission is bright in the
Lobe and the small region in the northeast shell (region 41),
which has a relatively low foreground absorption of \nh = 0.80 
(0.70 -- 0.94) $\times 10^{22}$~cm$^{-2}$.

\citet{2009AJ....137..354K} measured the proper motions of the AXP 1E\,2259+586
using \chandra\ ACIS-S observations in 2000 and 2006 pointed at the AXP.
The suggested initial position at which the supernova
occured and the AXP was born is north of the AXP and might even have been
in the area coinciding with the CO arm, depending on the real age of the SNR.
On the other hand, new near-infrared observations of the counterpart of
the AXP with the Keck II telescope using laser guide star adaptive optics
have been performed to measure the proper motion of the AXP. 
These data indicate that the initial position of the AXP was closer to 
the position of X-ray Lobe, approximately half way
between today's position of the AXP and the Si clump seen in blue
in Fig.\,\ref{softMgSi} (Tendulkar et al., in prep.).
Both results show that the position of the initial supernova explosion
was located west to northwest of the Lobe and suggest
that the interaction of the blast wave with a dense molecular cloud
produced the X-ray Lobe and, at the same time, caused the formation of
the reverse shock, which then ran into ejecta enhanced in Si or Fe. 
However, one has to keep in mind that the alignment of the Si enhanced region 
and the Lobe might well be a projection effect.

\section{Conclusions}

Through the analysis of the deep high-resolution data of the Lobe and the 
northeastern part of the SNR \ctb\ taken with \chandra\ ACIS-I, we confirm
ejecta emission inside the remnant. There is an enhancement in element 
abundances especially in the X-ray bright Lobe. 
We have thus unambiguously detected ejecta emission from an SNR associated 
with an AXP.
The ISM and ejecta emission indicate an age of $\sim$10 -- 20~kyr for the 
shocked X-ray emitting plasma. 

As one can see in the \xmm\ intensity image \citep[Fig.\,1,][]{2004ApJ...617..322S}
the SNR shell is brighter in the north and the south, while it is darker with 
diffuse structures in the outer parts of the same sector in which the Lobe 
is located (east-northeast). 
The analysis of \chandra\ data has shown that the foreground absorption \nh\ is 
lower in and around the Lobe (Fig.\ref{ftestpar}, lower panels), 
while the Fe L-shell emission is enhanced in the Lobe, with one smaller region
with significantly enhanced Si abundance.
The bright feature seen as the Lobe can be explained by the 
following scenario: the blast wave hit a dense cloud, evaporating and maybe 
encompassing it. 
Due to the interacton with the cloud, a reverse shock was formed and 
propagated towards the centre heating the ejecta.
The evaporated cloud is now visible as the Lobe with stronger X-ray emission.

\citet{2012ApJ...756...88C} report the detection of a GeV source with
\fermi\ at the position of \ctb. The \fermi\ source is located southwest of the
Lobe where the SNR appears dark, and thus most likely highly absorbed in X-rays.
Assuming that the $\gamma$-rays are produced in $\pi^0$-decays caused
by the interaction of the SNR shock with a dense interstellar cloud now visible
as the Lobe, \citet{2012ApJ...756...88C} derive a density of $\sim120$~cm$^{-3}$
for the cloud.  They point out the discrepancy between this value and the
density of 0.9~cm$^{-3}$, which we calculated from the X-ray data 
\citep{2006ApJ...642L.149S}.
However, as the density of the Lobe of 0.9~cm$^{-3}$ refers to the hot shocked 
X-ray emitting gas, which is likely to have a lower density than the original cold 
cloud, the new result from \citet{2012ApJ...756...88C} is consistent with, rather 
than incompatible, with our results.
As we have pointed out \citep{2006ApJ...642L.149S} and confirmed in the new
spectral analysis, the foreground absorption south of the Lobe (regions 38 and 51 
in this work) is high with \nh\ = $1.3 (1.1 - 1.5) \times 10^{22}$~cm$^{-2}$
and comparable to that in the CO arm. Therefore, our new \chandra\ study as well
as the new \fermi\ results by \citet{2012ApJ...756...88C} confirm that there was
and still is interaction between the shock wave of \ctb\ with the ambient dense
material. The remainder of this interaction is still visible as CO clouds 
presented in \citet{2006ApJ...642L.149S}.

\begin{acknowledgements}
M.S.\ acknowledges support by the Deutsche Forschungsgemeinschaft through the Emmy 
Noether Research Grant SA 2131/1.
P.P.P.\ and T.J.G.\ acknowledge support from NASA contract NAS8-03060.
\end{acknowledgements}

\bibliographystyle{aa} 
\bibliography{../../bibtex/ctb109,../../bibtex/xraytel,../../bibtex/my,../../bibtex/ism,../../bibtex/snrs}


\end{document}